\documentstyle[aps,epsfig]{revtex}

\def\nn{\nonumber}
\newcommand{\be}{\begin{equation}}
\newcommand{\ee}{\end{equation}}
\newcommand{\bea}{\begin{array}{c}}
\newcommand{\eaa}{\end{array}}
\newcommand{\ba}{\begin{eqnarray}}
\newcommand{\ea}{\end{eqnarray}}

\begin{document}
\date{\today}
\title{
Hard-thermal-loop resummed pressure of a degenerate \\
quark-gluon plasma}
\author{Rudolf Baier$^1$ and Krzysztof Redlich$^{1,2}$}
\address{$^1$
Fakult\"at f\"ur  Physik,  Universit\"at Bielefeld, \\
Postfach 10 01 31 , D-33501 Bielefeld, Germany\\
$^2$Institute for Theoretical Physics, University of Wroc\l aw,
PL-50204  Wroc\l aw, Poland}
\maketitle
\begin{abstract}
We compute the pressure of a finite density quark-gluon plasma
at zero temperature to leading order in hard-thermal-loop
perturbation theory, which includes the fermionic excitations
and Landau damping. The result is compared with the weak-coupling
expansion for finite  positive chemical potential
$\mu $ through order $\alpha_s^2$
and with a quasiparticle model with a mass depending on $\mu$.
\end{abstract}
\setcounter{section}{1}

\section*{}

In this letter we extend the recent work by Andersen, Braaten
and Strickland   \cite{andersen} by  applying hard-thermal-loop
(HTL) perturbation theory at the leading order \cite{pisarski}
to a relativistic quark-gluon plasma at
finite and positive chemical potential
           and at zero temperature.
These investigations are motivated by the observation of
the poor convergence properties of the perturbative
expansion of thermodynamic functions describing the quark-gluon
plasma at high temperature \cite{arnold}
and at high density \cite{freedman,kapusta}.
However,
 efficient resummation schemes have been recently developed in the
case of scalar field theory \cite{karsch} and of gauge
theories like QCD \cite{andersen,blaizot}, which
improve considerably the convergence  of the free energy at high
temperature.
%%%%%%%%%%%%%%%%%%%%%%%%%%%%%%%%%%%%%%%%%%%%%%%%%%%%%%%%%%

In the calculation of the high temperature  behavior of the pressure
of a gluon gas
     \cite{andersen}
the inverse HTL-resummed gluon propagator is important.
In the case of a degenerate quark-gluon gas at $T=0$
and non-vanishing chemical potential $\mu$ the improved
quark propagator plays the central role.
The ideal gas behaviour at $T=0$ and $\mu \neq 0$ is
 determined by a quark
loop

\begin{equation}
P_{ideal}(\mu ,T=0)=\lim_{T\to 0} {T\over V}{\rm Tr}\log {\rlap/p}
={{N_c}\over {12\pi^2}}\mu^4,
\label{eq1}
\end{equation}
where we state the result for a one-flavor ($N_f=1$) massless
quark-gluon gas with $N_c=3$ colors. The fermionic
Matsubara frequencies are given by $p_0=i(2n+1)\pi T +\mu$.
The $T=0$ limit is taken after the Matsubara sum
is performed.

The HTL-pressure $P_{\rm HTL}$ receives contributions from
three terms: $P_{\rm HTL}=P_{q}+P_{g}+\Delta P$, that is
from the HTL-resummed quark propagator, from the gluon propagator
and from possible counterterms, which  renormalize the pressure
\cite{andersen}.
The quark-loop contribution $P_q$ is obtained by
replacing the momentum
${\rlap/p}$
 in (\ref{eq1}) by ${\rlap/p}-\Sigma_{\rm HTL}$,
where $\Sigma_{\rm HTL}$ is the quark-selfenergy in the HTL
approximation \cite{klimov}. In the following we use the
expressions and notations as given in \cite{le}, i.e.

\begin{equation}
{\rlap/p}-\Sigma_{\rm HTL}=A_0\gamma_0-A_s\vec\gamma {\hat p} ~~ {\rm with}~~
A_0=p_0-{{m_f^2}\over p}Q_0({{p_0}\over p})~~{\rm and}~~
A_s=p_0+{{m_f^2}\over p}(1-{{p_0}\over p}Q_0({{p_0}\over p})),
\label{eq2}
\end{equation}
where $Q_0$ denotes the Legendre function. The quark thermal
mass at $T=0$ and  $\mu >0$ is given by

\begin{equation}
m_f^2={{N_c^2-1}\over {4N_c}} {{\alpha_s (\mu_4)}\over \pi} \mu^2,
\label{mfq}
\end{equation}
with the $\overline{MS}$ renormalization scale $\mu_4$.
Therefore, for $N_c=3$ we have $m_f<\mu$.
In terms of the functions $A_0,A_s$ the pressure is

\begin{equation}
P_q=2N_c
 {\sum\hspace*{-0.5cm}{\int}}
   \log (A_s^2-A_0^2),
\label{eq3}
\end{equation}
where
${\sum\hspace*{-0.37cm}{\int}}
 \equiv T\sum_n\mu_3^{3-d}\int {{d^dp}\over {(2\pi)^d}}$
represents the sum-integral in $d$-dimension, $\mu_3$ denotes the
$MS$ renormalization scale,
necessary for the renormalization of the pressure
using dimensional regularization  for
 $d\neq 3$ dimensions      \cite{andersen}.

In order to evaluate (\ref{eq3}), it is useful to use
 the spectral representation  of the quark-propagator.
 Having in mind the linear dependence of $A_0$ and $A_s$ on
 $m_f^2$ we first separate the ideal gas contribution by

\begin{equation}
 P_q-P_{ideal}=
 \int_0^{m_f^2}dm^2 {{\partial P_q}\over {\partial m^2}}
 =
 2N_c\int_{m_f^2}^0 {{dm^2}\over {m^2}}
 {\sum\hspace*{-0.5cm}{\int}}
      \{ (p_0-p)\Delta_+(p_0,p;m^2)+
         (p_0+p)\Delta_-(p_0,p;m^2) \}.
\label{eq4}
\end{equation}
For $\Delta_{\pm}(p_0,p;m^2)=(A_0\mp A_s)^{-1}$ the following
dispersion relations hold

\begin{equation}
\Delta_\pm (p_0+i\eta ,p;m^2)=
-\int_{-\infty}^{+\infty } {{dq_0}\over {2\pi}}
{{\rho_\pm (q_0,p;m^2)}\over {q_0-p_0-i\eta }}.
\label{eq5}
\end{equation}
The sum rules for the HTL spectral function

\begin{equation}
\int_{-\infty}^{+\infty } {{dp_0}\over {2\pi}}
(p_0\mp p)\rho_\pm (p_0,p;m^2)=0
\label{eq6}
\end{equation}
turn out to be useful in our further discussion.
Using (\ref{eq5}) and (\ref{eq6}) we find for the expression (\ref{eq4}),

\begin{equation}
P_q=P_{ideal} + 2N_c \int_0^{m_f^2}
{{dm^2}\over {m^2}}
\int_p \int_\mu^\infty {{dp_0}\over {2\pi}}
\{
(p_0-p)\rho_+(p_0,p;m^2) +
(p_0+p)\rho_-(p_0,p;m^2)
\}.
\label{eq7}
\end{equation}
The derivation of (\ref{eq7}) is based on the observation that the
expression in the curly brackets of (\ref{eq4}) is
a meromorphic function in $p_0$, which decreases faster than
$p_0^{-1}$ for $|{p_0}|  \to \infty$, because of (\ref{eq6}).
This allows to perform the Matsubara frequency sum and
to take the $T=0$ limit in the standard way      \cite{le}.

The representation (\ref{eq7}) covers the following contributions:
(i) the fermion quasi-particle excitations for $p_0>p$,
which are solutions of ${\rm Re}\Delta_\pm^{-1} (p_0=\omega_\pm,p)=0$.
In (\ref{eq7}) they contribute for $\omega_\pm(p)\geq \mu$.
(ii) the Landau damping contribution restricted by $\mu\leq p_0\leq p$.

In order to proceed we state the
 explicit expressions for the spectral functions,

\ba
   \rho_\pm (p_0,p;m^2) &=&
 2\pi \left[
 Z_\pm(p) \delta (p_0-\omega_\pm(p)) +
 Z_\mp(p) \delta (p_0+\omega_\mp(p)) \right]
+{{\pi}\over p} m^2 (1\mp x)
   \theta (1-x^2)
\nn \\  &&  \cdot
\left[
(p(1\mp x) \pm
 {{m^2}\over {2p}} [(1\mp x)L\pm 2 ])^2
+
{{\pi m^4}\over {4p^2}} (1\mp x)^2 \right]^{-1},
\label{eq8}
\ea
with $x=p_0/p$, $L=\log    |{{x+1}\over {x-1}}|$ and the residues
at the quasi-particle poles $Z_\pm(p)={{\omega_\pm^2(p)-p^2}\over
{2m^2}}$.
Next we consider the contributions (i) and (ii) separately and
decompose
\be
 P_q=P_{ideal}+P_{q,quasi}+P_{q,Ld}+P_{q,sub}.
\ee
The quasi-particle contribution $P_{q,quasi}$ to the pressure $P_q$ is
obtained as follows. Due to the $\delta $-functions in (\ref{eq8})
the $p_0$-integration in (\ref{eq7}) is straightforward.
We also note   the identities
$
d\omega_\pm=(\omega_\pm \mp p)
Z_\pm (p) {{dm^2}\over {m^2}}$, which  allow to perform the $m^2$-integration.
The quark $(+)$ mode  contribution has an ultraviolet divergence for $d$=3
dimensions. Following the procedure  of     \cite{andersen}
this divergence is isolated by subtracting expressions from
the integrands to obtain finite results for $d=3$, and evaluating
the subtracted integrals in $d=3-2\epsilon ,$  ($\epsilon >0$)
dimensions.  The subtracted terms take into account that
$ \omega_+\to p+{{m_f^2}\over p} -
{{m_f^4}\over {2p^3}}\log {{2p^2}\over {m_f^2}}$,
for $p>>m_f$      \cite{klimov}.
The quark $(-)$  mode contribution is ultraviolet finite, therefore
it does not require subtracted terms. The finite
quasi-particle contribution for $d=3$ reads

\ba
  P_{q,quasi}=
2N_c \int {{d^3p}\over {(2\pi )^3}}&&  \{
%\left[
(\omega_+(p)-\mu )\theta (\omega_+-\mu )\theta (\mu -p) +
(\omega_-(p)-\mu )\theta (\omega_--\mu )\theta (\mu -p)
\nn \\ &&   +
(\omega_-(p)-p )\theta (p-\mu )
 -
(\omega_+(p)-p )\theta (\mu -p) +
[ \omega_+(p)
\nn \\ &&
 -\sqrt {p^2+2m_f^2} +
{{m_f^4}\over {2(p^2+2m_f^2)^{3/2}}} (\log {
{{2(p^2+2m_f^2)}\over {m_f^2}}} -1)] \} .
%\right].
\label{eq9}
\ea
The Landau damping term $P_{q,Ld}$ is treated in an
analogous way. The dependence of $\rho_\pm$ on $m^2$
allows to perform the $m^2$-integration in (\ref{eq7}) explicitly.
The ultraviolet divergence, which is also present in this
contribution, is treated as described above. The finite term for
$d=3$ becomes

\ba
P_{q,Ld}=
-4N_c   \int   {{d^3p}\over {(2\pi )^3}}
\theta (p-\mu )  & \int_\mu^p & {{dp_0}\over {2\pi}}
     [
\Phi_+(p_0,p;m_f^2) -
\Phi_-(p_0,p;m_f^2)
 \nn \\ &&
 +{{\pi m_f^4p_0}\over {p^3}} {{1}\over {(p^2-p_0^2+ 2 m_f^2)}}
+{{\pi m_f^4}\over {2p^4}}
\log { ({{p+p_0}\over {p-p_0}}) }
],
\label{eq10}
\ea
where the angles $\Phi_\pm$ are given by

\begin{equation}
\Phi_\pm (p_0,p;m_f^2)=
\arctan
{{\pi (1\mp x)m_f^2}\over
{2p[p(1\mp x)+ {{m_f^2}\over {2p}} (2\pm (1\mp x)L)]}}.
\label{eq11}
\end{equation}
The terms in (\ref{eq10}) which are subtracted in order to make it
finite are constructed by observing that
$\Phi_+-\Phi_- \to
 - {{\pi m_f^4}\over {2p^4}}
[{{2x}/(1 -x^2 + 2m_f^2/p^2)} +L]
$
  for $p>>m_f$.

Finally, we discuss the term $P_{q,sub}$, which
contains the renormalized subtracted contributions
evaluated in the limit $\epsilon \to 0$. We note
that the quasi-particle and Landau damping terms
in (\ref{eq7}) have poles proportional to
$1/\epsilon^2$ and to $1/\epsilon$, which however cancel when adding these
two contributions. Therefore no counter term is required:

\begin{equation}
 \Delta P_q= 0.
\end{equation}
 The remaining finite term reads

\begin{equation}
P_{q,sub}=
N_c {{m_f^4}\over {4\pi^2}}
[ -{1\over 2} (\log {{{m_f^2}\over {2\mu^2}}})^2
-4\log 2 +{5\over 2}
- \int_0^{-\log {{{2 m_f^2}\over {\mu^2}}}}
{{tdt}\over {e^t-1}}
~].
\label{eq12}
\end{equation}
We observe  from (\ref{eq12}) that $P_q$ includes terms of
$O(m_f^4 {\hat =} \alpha_s^2)$. Consequently we have to consider
contributions of $O(m_g^4 {\hat =} \alpha_s^2)$ arising
from the HTL-resummed gluon loop.
In leading order the result is already given in      \cite{andersen}
when discussing the $T=0$ limit of the pressure, namely by

\begin{equation}
P_g=-(N_c^2-1)
{9\over {64\pi^2}}m_g^4
(\log {{{m_g}\over {\bar {\mu_3}}}} -0.332837),
\label{eq13}
\end{equation}
where the $\overline {MS}$ scale ${{\bar {\mu_3}}}^2=4\pi e^{-\gamma }\mu_3^2$
is introduced.
However,
here we have to replace the thermal gluon mass, which is
$
m_g^2={{4 \pi \alpha_s}\over {3}}  T^2
$
in a hot gluon gas, by
$
m_g^2={2\over 3}{{\alpha_s}\over \pi}\mu^2
$
\cite{toimela} for the case of degenerate quark-gluon gas.
We note that $m_g$ coincides with $m_f$.
In the following we fix
the renormalization scale $\bar {\mu_3}$ by
$\bar {\mu_3}=0.717 m_f$, such that $P_g=0$, and only the
quark-loop contributes to $P_{\rm HTL}$ (besides
the proper counter term for the gluon-loop).

For later comparisons with the weak-coupling expansion we
expand $P_q$ in powers of $m_f/\mu <<1$, and
we find

\begin{equation}
{{P_q}\over {P_{ideal}}}\simeq
1-2N_c {{m_f^2}\over {\mu^2}}
+O({{m_f^4} \over {\mu^4}}) =1-4{{\alpha_s}\over \pi }+ O(\alpha_s^2).
\label{eq14}
\end{equation}
We note that the leading correction of $O(m_f^2/\mu^2)$ is
exclusively due to the quark (+) mode contained in
$P_{q,quasi}$ of (\ref{eq9}).

The pressure of a degenerate quark-gluon plasma
is calculated through order $\alpha_s^2$ in the
weak-coupling expansion \cite{freedman,kapusta}.
For one flavour $(N_f=1)$ it is given by

\begin{equation}
P_{QCD}=P_{ideal}
\left[
1-2{{\alpha_s(\mu_4)}\over {\pi}}
-(9.267+
\log {({{\alpha_s(\mu_4)}\over \pi})}
-{{31}\over {3}}
\log {{\mu\over {\mu_4}}})
({{\alpha_s(\mu_4)}\over \pi})^2
\right]
\label{eq15}
\end{equation}
with
$
{{\alpha_s(\mu_4)}\over \pi}=
{{12}\over {31\bar L}}
(1-{{804}\over {961}}
{{\log {\bar L}}\over {\bar L}})
$
for $N_c=3$ and $\bar L =\log {(\mu_4^2/\Lambda^2_{\overline {MS}})}$.

Although $P_{QCD}$ in (\ref{eq15}) depends on
 $\mu_4$ beginning at order $\alpha_s^3$,
the dependence on $\mu_4$ turns out to be
rather strong. For $\mu /\Lambda_{\overline {MS}} = 2$,
$P_{QCD}/P_{ideal}$ varies from 0.68 to 0.82
under the variation from $\mu_4= 2 \mu$   to $\mu_4= 8 \mu$.
 With increasing $\mu /\Lambda_{\overline {MS}}$ the variation
obviously decreases, and consequently the
difference between the
$O(\alpha_s)$ and
$O(\alpha_s^2)$ contributions in (\ref{eq15}) is diminished.
This is shown in Fig.1 for the choice $\mu_4=4\mu$.

When comparing the
$O(\alpha_s)$  correction term of $P_{QCD}$ in (\ref{eq15})
with the one of the HTL-resummed pressure $P_{\rm HTL}$
at $T=0$ (\ref{eq14}) we note a difference in magnitude by a factor 2.
As in the case of the gluon gas at high $T$      \cite{andersen}
$\rm HTL$ resummation overestimates the
$O(\alpha_s)$  term. In order to correct for this difference,
we add a correction term to $P_{\rm HTL}$,
namely $P_{corr}=2{{\alpha_s}\over \pi} P_{ideal}$.
We define the pressure

\be
 P\equiv P_{\rm HTL}+P_{corr} ,
 \label{eqn}
 \ee
and attribute $P_{corr}$ to quasi-particle
interactions  \cite{andersen}. Presently, however,
we are only able to approximate the
corresponding series  for $P_{corr}$
by its leading $O(\alpha_s)$  contribution.

%%%%%%%%%%%%%%%%%%%%%%%fig 1
\begin{figure}[htb]
\begin{center}
\epsfig{file=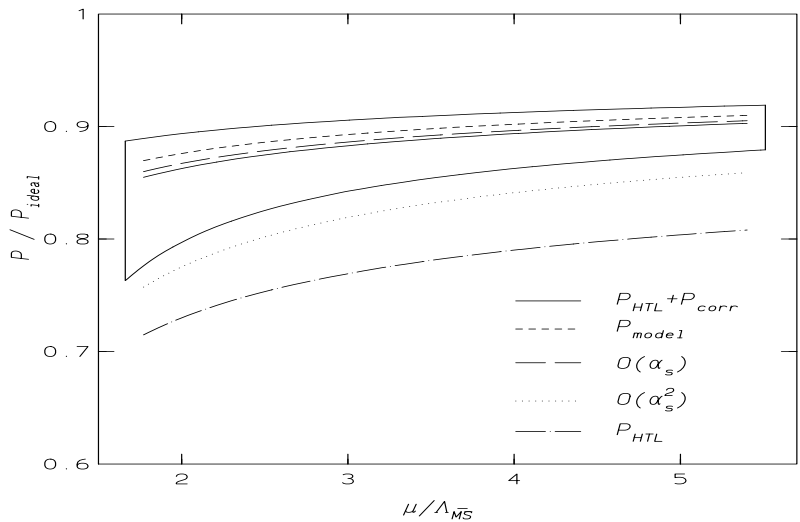, width=110mm,height=95mm}
\end{center}
\vspace*{-0.8 cm}
\caption
      {{\rm
The pressure for a degenerate quark-gluon plasma normalized to the
ideal gas result   as a function of $\mu /\Lambda_{\overline {MS}} \,$.
The $P_{HTL}+P_{corr}$
pressure is shown for  $\mu_4 =4 \mu$ and
as a band that corresponds to varying $\mu_4$ from
$\mu_4 =2 \mu$ to $\mu_4 =8 \mu$ (solid line).
The weak-coupling expansions through order $O(\alpha_s (\mu_4))$
and $O(\alpha_s^2(\mu_4))$ are shown by long-dashed
 and dotted lines correspondingly;
the HTL pressure $P_{HTL}$  is indicated by the
dashed-dotted line, and the pressure $P_{model}$ for non-interacting
massive, $m=m_f (\mu_4)$, fermionic quasi-particles is shown as
short-dashed  line.
All are calculated for $\mu_4=4 \mu$. }}
\end{figure}
%%%%%%%%%%%%    INPUT FIGURES   %%%%%%%%%%%%%%%%%%%%%%%%%%

In Fig.1 we show $P_{\rm HTL}$ as a function of $\mu /\Lambda_{\overline {MS}}$
evaluated for $\mu_4=4\mu$, indicating the overestimate of the
$O(\alpha_s)$ correction. Numerically the main contribution
to $P_{\rm HTL}$ is due to the quark mode $\omega_+(p)$ in (\ref{eq9}),
which is arising from the momentum interval $\mu \geq p > 0$.
The result for $P$ is plotted in Fig.1  for $\mu_4=4\mu$ and
as a band which
corresponds to the variation of the scale
$\mu_4$ by $2\mu\leq \mu_4 \leq 8\mu$.
We observe  that the
$O(\alpha_s^2)$  corrections of the weak-coupling
expansion (\ref{eq15}) lead to a much bigger deviation from the
ideal gas behaviour than the  HTL-resummed $P$, which
follows more closely the
$O(\alpha_s  )$ behaviour of (\ref{eq15}) in the considered region of
$\mu /\Lambda_{\overline {MS}}$.

We can not compare at present our result for the
pressure at $T=0$, $\mu >0$, with lattice results, as it could
be done  at $\mu =0$ and $T>0$. Instead we evaluate
$P$ for non-interacting massive, $m\neq 0$,
fermionic quasi-particles $(N_f=1, N_c=3)$ with finite
chemical potential. The corresponding pressure reads
\cite{cleymans}

\begin{equation}
P_{model} = {{m^4}\over {4\pi^2}}
\left[
{\mu\over m} ({{\mu^2}\over {m^2}}-1)^{1\over 2}
({{\mu^2}\over {m^2}}-{5\over 2}) +
{3\over 2}
\log {[ {\mu\over m}+
\sqrt {{{\mu^2}\over {m^2}}-1}]}
\right],
\end{equation}
with $\mu >m$. Choosing $m=m_f$ of (\ref{mfq}) we realize that
through
$O(\alpha_s  )$  $P_{model}$ coincides with
$P_{QCD}$ (\ref{eq15}) as well as  with $P$ (\ref{eqn}). This is clearly
 seen in Fig.1 where $P_{model}$ is plotted for $\mu_4=4\mu$.
Indeed the quasi-particle approximation \cite{satz}
turns out to be an excellent effective description of
the degenerate quark-gluon plasma at $T=0$, when the quasi-particle
mass $m$ is identified with the quark thermal mass $m_f$
of the  HTL  approach. This agreement
is also shown  in  Fig.2, where we summarize  different
approximations for the pressure discussed in this letter,
which we normalize to the ideal gas for $T=0$, $m=0$ and positive
$\mu$.
For the choice $\mu_4=4\mu$ we plot in Fig.2 the ratio
$P/P_{ideal}$ as a function of the coupling $\alpha_s (\mu_4)$
in the region $0  \leq \alpha_s \leq 0.4$ for
$P_{\rm HTL}$, $P=P_{\rm HTL}+P_{corr}$, $P_{QCD}$ through
$O(\alpha_s  )$ and
$O(\alpha_s^2)$, and for $P_{model}$.

Pisarski and Rischke in a recent paper \cite{rischke} conjecture
that perturbation theory at $T=0, \mu >0$ may work better than for
$T >0$, i.e. they estimate its breakdown when the coupling
becomes as large as $\alpha_s \sim 2.4$. From Fig.2
one may estimate a "critical" $\alpha_s \sim 0.4$, i.e. where the
$\alpha_s^2$ term becomes as big as the $\alpha_s$ term of $P_{QCD}$.
At $T > 0$ a corresponding estimate gives $\alpha_s \sim 0.1 - 0.2$.

Only after the inclusion of, at least,  the  next-to-leading order correction
in the HTL-perturbation scheme one could judge if
the prediction for the pressure of a degenerate quark-gluon
plasma becomes considerably more stable,
especially when compared to the weak-coupling expansion (\ref{eq15}).
 Then it would be extremely interesting \cite{wilczek} to compare with
numerical   lattice results for
the $T=0$ finite density plasma which also, however,
 still  have  to be worked out.

%%%%%%%%%%%%%%%%%%%%%%%fig 2
\begin{figure}[htb]
\begin{center}
\epsfig{file=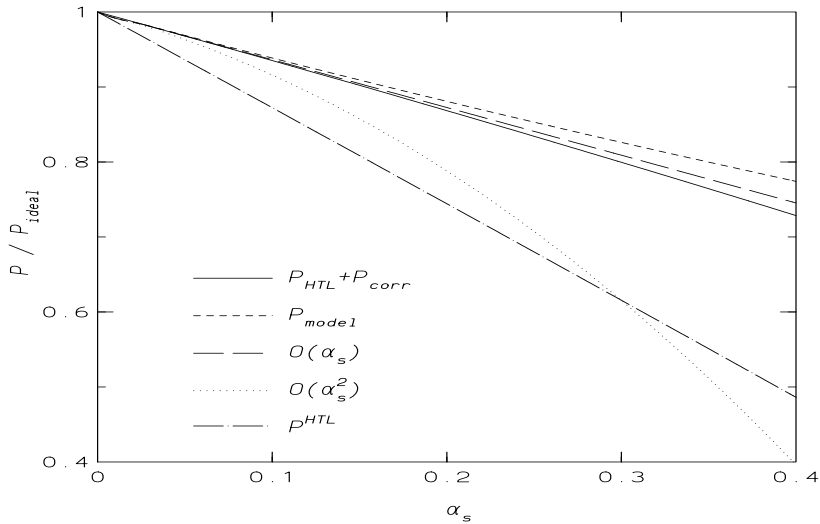, width=110mm,height=95mm}
\end{center}
\vspace*{-0.9 cm}
\caption
      {{\rm
The pressure for a degenerate quark-gluon plasma normalized to the
ideal gas result   as a function of $\alpha_s (\mu_4)$. The contributions
to the pressure correspond  to those of Fig.1, consistently calculated
 with $\mu_4=4 \mu$. }}
\end{figure}
%%%%%%%%%%%%%%%%%%%%%%%%%%%%%%%%%%%%%%%%%%%%%%%%%%

After having completed this letter we have found a  relevant new paper by
Andersen, Braaten, and    Strickland \cite{l}
 where the quark contribution to the  free energy
of a hot ($T\neq 0$) but baryon free ($\mu =0$) quark-gluon plasma
is calculated to leading order in HTL-perturbation theory.
In another recent paper \cite{areb} the calculation of the baryon density
in a  self-consistent approximation in the sense of \cite{blaizot,baym}
is presented as a function of
$\mu /\Lambda_{\overline {MS}} \,$.

%\section*{}
%
This work is supported in part by
 Deutsche Forschungsgemeinschaft (DFG project
 KA 1198/4-1).
We thank H.~Satz for interesting discussions and A.~Rebhan
 for useful comments.
We especially wish to thank J.~O.~Andersen for pointing out to us
a mistake in Eqs. (\ref{eq10}) and (\ref{eq12}) of the early
version of this paper.
One of us (K.R.) also
   acknowledges stimulating discussions with F.~Karsch and D.~E.~Miller.

\end{document}